\documentstyle[12pt,particles,journals,symbols,rotate,epsfig]{article}
\begin{document}
\setlength{\baselineskip} {2.5ex}
\hyphenation {che-ren-kov}
\hyphenation {non-per-tur-ba-tive}
\hyphenation {per-tur-ba-tive}
\begin {center} 
{\bf Submitted to the Proceedings,\\  
XXXVII International Winter Meeting on Nuclear Physics\\
Bormio, Italy, Tel Aviv U. Preprint TAUP-2562-99, Jan. 1999.\\}
%Tel Aviv U. Preprint TAUP 2207-94\\
%Bulletin Board: hep-ph@xxx.lanl.gov/9410215}
\end{center}
\vspace{0.3cm}
\begin{center}
{\Large \bf  Hadron-Photon Interactions in COMPASS \\}
\vspace{0.3cm}
\end {center}
\vspace{0.3cm}
\begin {center}
{\bf
Murray A. Moinester, Victor Steiner \\
Raymond and Beverly Sackler Faculty of Exact
Sciences,\\
School of Physics,
Tel Aviv University, 69978 Ramat Aviv, Israel\\ 
E-mail: murraym@tauphy.tau.ac.il,~steiner@lepton.tau.ac.il\\
\vspace{0.3cm}
Serguei Prakhov\\
Joint Institute for Nuclear Research,\\
Dubna, 141980 Moscow Region, Russia\\
E-mail:  prakhov@nusun.jinr.dubna.su\\}
\end {center}
\vspace{1cm}
{\bf Abstract:\\}

\noindent
The CERN COMPASS experiment will 
investigate hadron-photon interactions, 
to achieve a unique Primakoff Coulomb physics program based on
hadron polarizability, hybrid mesons, chiral anomaly, and meson radiative
transition studies. In COMPASS, pion (kaon) polarizabilities and radiative
transitions will be measured via dedicated Primakoff effect reactions such
as $\pi^- \gamma \rightarrow {\pi^-}' \gamma$ and $\pi^- \gamma
\rightarrow M$~(meson). We will also study pionic and kaonic hybrid
mesons
in COMPASS by Primakoff production.  We will also study the radiative
transition of a pion to a low mass two-pion system, $\pi^- \gamma
\rightarrow {\pi^-}' \pi^0$, for a measure of the chiral anomaly amplitude
F$_{3\pi}$ (characterizing $\gamma \rightarrow 3 \pi$). Our objectives are
to determine new properties of hadrons and to provide important new tests
of QCD chiral dynamics.\\

\noindent
The CERN COMPASS experiment use 50-280 GeV beams ($\mu$,
$\pi$, K, p)  and a virtual photon target, and magnetic
spectrometers and calorimeters to measure the complete kinematics of
hadron-photon reactions.  
The COMPASS experiment is currently under construction, and
scheduled in 2000 to begin data runs, with muons initially and hadrons
following.  The Primakoff program in COMPASS is approved as the first
hadron physics program.  We carry out simulation studies to optimize the
beam, detector, trigger, and hardware/software for achieving high
statistics data with low systematic uncertainties in this program.
For kaon studies, our results will be the first ever. For pion
studies, we will improve previous results by more than two orders of
magnitude. 
We will implement special detectors and triggers
for hadron-photon reactions.  
We will prepare
the COMPASS hadron-photon Primakoff studies by
setting up
with muon-photon Primakoff tests.

\begin{center}
{\bf  1. Hadron-Photon Interactions:\\}
\end{center}

\noindent
The COMPASS physics programs \cite {selex,compass} include
studies of
hadron-photon  Primakoff interactions using 50-280 GeV/c negative beams
(pions, kaons) and positive beams (pions, kaons, protons)  together with a
virtual photon target in dedicated data runs. 
Pion and kaon and proton polarizabilities, hybrid mesons, the
chiral anomaly, and radiative transitions can be studied in this way, and
can provide significant tests of QCD and chiral perturbation theory
($\chi$PT)  predictions. 
Preliminary COMPASS Primakoff planning studies appear in detail in 
the proceedings of the 1997 Mainz Chiral Dynamics workshop and Prague
COMPASS summer school
\cite {cd2,hadron1}. 
The objectives and significance of the studies 
are further described below. 

\vspace{.5cm}
\noindent
{\bf Pion Polarizabilities\\}
%\vspace{.5cm}

\noindent
For the pion polarizability, $\gamma\pi$ scattering was measured (with
large uncertainties)  with 40 GeV pions \cite{anti1} via radiative pion
scattering (pion Bremsstrahlung) in the nuclear Coulomb field: 
\begin{equation}
\label{eq:polariz}
\pi + Z \rightarrow \pi' + \gamma + Z'.
\end{equation}
In this measurement, the incident pion Compton scatters from a virtual
photon in the Coulomb field of a nucleus of atomic number Z; and the final
state $\gamma$ and pion are detected in coincidence.  The radiative pion
scattering reaction is equivalent to $\gamma$ + $\pi^{-}$ $\rightarrow$
$\gamma$ + $\pi^{-}$ scattering for laboratory $\gamma$'s of order 1 GeV
incident on a target $\pi^{-}$ at rest. It is an example of the well
tested Primakoff formalism \cite{jens,ziel2} that relates processes
involving real photon interactions to production cross sections involving
the exchange of virtual photons. 

In the 40 GeV radiative pion scattering experiments, it was shown
experimentally \cite{anti1} and theoretically \cite{galp} that the
Coulomb amplitude clearly dominates, and yields sharp peaks in
t-distributions at very small squared four momentum transfers (t) to the
target nucleus t $\leq 6 \times 10^{-4}$ (GeV/c)$^{2}$. Backgrounds from
strong processes were low. The backgrounds are expected to be lower at the
higher energies planned for the CERN COMPASS experiment. 

For the $\gamma\pi$ interaction at low energy, $\chi$PT provides a
rigorous way to make predictions via a Chiral Lagrangian written in terms
of renormalized coupling constants L$^r_i$ \cite{gass1}. With a
perturbative expansion of the effective Lagrangian, the method establishes
relationships between different processes in terms of the L$^r_i$. For
example, the radiative pion beta decay and electric pion polarizability
are expressed as \cite{hols1}:
\begin{equation}
F_A/F_V = 32\pi^2(L^r_9+L^r_{10});~ \bar{\alpha}_{\pi} = 
\frac{4\alpha_f}{m_{\pi}f^{2}_{\pi}}(L^r_9+L^r_{10});
\label{eq:fafv}
\end{equation}
\noindent
where f$_\pi$ is the pion decay constant, m$_\pi$ is the pion mass, 
F$_A$ and F$_V$ are the axial
vector and vector coupling constants in the decay, and $\alpha_f$ is the
fine structure constant. The experimental ratio F$_A$/F$_V$ = 0.45 $\pm$
0.06, leads to $\bar{\alpha}_{\pi}$ = -$\bar{\beta}_{\pi}$ = 2.7 $\pm$
0.4, where the error shown is due to the uncertainty in the F$_A$/F$_V$
measurement \cite{hols1,babu2}.  All polarizabilities 
are expressed in Gaussian units of $10^{-43}$ cm$^3$. The $\chi$PT
prediction \cite {hols1} for the pion polarizability is
$\bar{\alpha}_{\pi}$ = 2.7. Ref. \cite {hols1} showed that meson exchange
via a
pole diagram involving the a$_1$(1260) resonance provides the main
contribution ($\bar{\alpha}_{\pi}$ = 2.6) to the polarizability.
Ref. \cite{xsb} assuming a$_1$ dominance finds $\bar{\alpha}_{\pi}$ = 1.8.
For
the kaon, the $\chi$PT polarizability prediction \cite {hols1} is
$\bar{\alpha}_{K}$ = 0.5. A more extensive theoretical study of kaon
polarizabilities was given recently \cite {ev}.

The pion polarizabilities deduced by Antipov et al. \cite{anti1} in their
low statistics
experiment ($\sim$ 7000 events) were $\bar{\alpha}_{\pi} =
-\bar{\beta}_{\pi} = 6.8 \pm 1.4 \pm 1.2$. It was assumed in the analysis
that $\bar{\alpha}_{\pi} + \bar{\beta}_{\pi} = 0$, as expected
theoretically \cite {hols1}.  The deduced polarizability value, ignoring
the large error bars, is about three times larger than the
$\chi$PT prediction. {\bf The available polarizability results have large
uncertainties. There is a clear need for new and improved radiative pion
scattering data.} The pion polarizabilty provides an important test of 
chiral dynamics.  

\newpage
\vspace{.5cm}
\noindent
{\bf Hybrid Mesons\\}
%\vspace{.5cm}

\noindent
The hybrid ($q\bar{q}g$) mesons, along with glueballs ($gg$) are one of
the most amazing consequences of the non-abelian nature of QCD. Detection
of these exotic states is a long-standing experimental puzzle. The most
popular approach for the hybrids search is to look for the "oddballs" -
mesons with the quantum numbers not allowed for the 
ordinary $q\bar{q}$ states, for
example J$^{PC}= 1^{-+}$, decaying to $\eta \pi$, $\eta ' \pi$, $f_1(1285)
\pi$, $b_1(1235) \pi$, etc. 

From more than a decade of experimental efforts at IHEP \cite
{ihep1,ihep2,ves}, CERN \cite {na12}, KEK \cite{kek} and BNL \cite {E852},
several hybrid candidates have been identified. The most recent
information came from BNL E852 experiment \cite {E852} which studied
the $\pi^- p$ interaction at 18 GeV/c. They reported J$^{PC}= 1^{-+}$
resonant
signals in $\eta \pi^-$ and $\eta \pi^0$ systems as well as in $\pi^+
\pi^- \pi^-$, $\pi^- \pi^0 \pi^0$, $\eta ' \pi^-$ and $f_1(1285) \pi^-$.
At the same time, a VES group \cite {ves} has published analysis of $\eta
\pi^-$, $\eta ' \pi^-$, $f_1(1285) \pi^-$, $b_1(1235) \pi^-$ and $\rho
\pi^-$ systems production in $\pi^- Be$ interaction at 37 GeV/c. Although
the J$^{PC}= 1^{-+}$ wave is clearly seen by VES in all channels, there is
no indication for the presence of a narrow ($\Gamma \sim 0.2~ GeV$)
resonance in any of them. But an observed abnormally high ratio of $\eta '
\pi$ to $\eta \pi$ P-wave is considered as evidence for the hybrid nature
of this exotic wave.  

It should be mentioned that the partial wave analysis (PWA) of systems
such as $\eta \pi$ or $\eta ' \pi$ in the mass region below 2 GeV is
particularly difficult. This is so because (1) this region is dominated by
the strong $2^+$ "background" (a1 resonance), and (2) that the PWA may
give ambiguous results \cite {ihep2} for the weaker $1^{-+}$ wave. The
problem is that the PWA of the $\eta\pi$ system must take into account S,
P and D waves, and the number of observables is not sufficient to solve
unambiguously all equations. Looking at the partial wave solutions as a
function of mass, each partial wave can have as many as eight different
curves to describe its strength and phase, as discussed in ref. \cite
{ihep2}. It is therefore extremely important to have extra information
from different hybrid production mechanisms where the physics is different
and such ambiguities may look different.  Only by comparing results of
different experiments in this way, can we establish unambiguously the
existence or non-existence of hybrid (or exotic) meson states.
COMPASS will look for Primakoff production of pionic and kaonic hybrid
mesons in the 1-2 GeV mass
region, including all hybrid candidates from previous studies.  

\vspace{.5cm}
\noindent
{\bf Chiral Axial Anomaly\\}
%\vspace{.5cm}

\noindent
The Chiral Axial Anomaly can also be studied with 50-280 GeV
pion beams. For the
$\gamma$-$\pi$ interaction, the O(p$^4$) chiral lagrangian \cite
{gass1,donn1} includes Wess-Zumino-Witten (WZW)  terms \cite{wzw,bij3},
which lead to a chiral anomaly term \cite{wzw,bij3,anti2} in the
divergence equations of the currents. This leads directly to interesting
predictions \cite{bij3} for the processes 
$\pi^0 \rightarrow 2 \gamma$ and
$\gamma \rightarrow 3 \pi$; and other processes as well \cite{bij3}. The
two processes listed are described by the amplitudes F$_{\pi}$ and
F$_{3\pi}$, respectively. 

The chiral anomaly term leads to a prediction for F$_{\pi}$ and F$_{3\pi}$
in terms of $N_c$, the number of colors in QCD; and f$_\pi$, the charged
pion
decay constant. The O(p$^4$) F$_{\pi}$ prediction 
for $\pi^0 \rightarrow 2 \gamma$ 
is in agreement with
experiment \cite {bij3}. The F$_{3\pi}$ prediction is \cite {ca,hols2}: 

\begin{eqnarray}
F_{3\pi} = {N_c (4 \pi \alpha)^{1 \over 2} 
\over 12 \pi^2 f_{\pi}^3} \sim 9.7 \pm 0.2 ~GeV^{-3},~ O(p^4). 
\label{eq:f3pi}
\end{eqnarray}
\noindent
The experimental confirmation of this equation would demonstrate that the
O(p$^4$) terms are sufficient to describe F$_{3\pi}$.

The amplitude F$_{3\pi}$ was measured by Antipov et al. \cite{anti2} at
Serpukhov with 40 GeV pions. Their study involved pion production by a pion in
the nuclear Coulomb field via the Primakoff reaction:

\begin{equation}
 \pi^- + Z \rightarrow {\pi^-}' + \pi^0 + Z'.
\label{eq:anomaly}
\end{equation}
\noindent
In the one-photon exchange domain, eq. \ref{eq:anomaly} is equivalent to: 

\begin{equation}
 {\pi^-} + \gamma  \rightarrow  {\pi^-}' + {\pi^0},          
\label{eq:eqanomaly}
\end{equation}
\noindent
and the 4-momentum of the virtual photon is k = $P_Z$-$P_{Z'}$. The cross
section formula for the Primakoff reaction depends on $F_{3\pi}^2$. The
Antipov et al. data sample (roughly 200 events) covering the ranges $-t <
2. \times 10^{-3} (GeV/c)^2$ and $s (\pi^-\pi^0) < 10.~m_{\pi}^2$. The
small t-range selects events predominantly associated with the exchange of
a virtual photon, for which the target nucleus acts as a spectator.
Diffractive production of the two-pion final state is blocked by G-parity
conservation. The experiment \cite{anti2} yielded F$_{3\pi}=12.9 \pm 0.9
(stat) \pm 0.5 (sys) ~GeV^{-3}$. This result differs from the O(p$^4$)
expectation by at least two standard deviations; so that the chiral
anomaly prediction at O(p$^4$)  is not confirmed by the available $\gamma
\rightarrow 3\pi$ data. 
                                 
Bijnens et al. \cite{bij3} studied higher order $\chi$PT corrections in
the abnormal intrinsic parity (anomalous) sector. They included one-loop
diagrams involving one vertex from the WZW term, and tree diagrams from the
O(p$^6$) lagrangian. They determine parameters of the lagrangian via vector
meson dominance (VMD) calculations. The higher order corrections are small for
F$_{\pi}$. For F$_{3\pi}$, they increase the lowest order value from 7\% to
12\%. The one-loop and  O(p$^6$) corrections to F$_{3\pi}$ are comparable in
strength. The loop corrections to F$_{3\pi}$ are not constant over the whole
phase space, due to dependences on the momenta of the 3 pions. The average
effect is roughly 10\%, which then increases the theoretical prediction by 1
GeV$^{-3}$. The prediction is then $F_{3\pi} \sim  10.7$, closer to the 
data. The limited accuracy of the existing data, together with the new
calculations of Bijnens et al., motivate an improved and more precise
experiment.
The expected number of near threshold two-pion events in COMPASS is
several orders of magnitude larger than in all previous experiments
\cite{ca}.
      
\vspace{.5cm}
\noindent
{\bf Radiative Transitions\\}
%\vspace{.5cm}

\noindent
Radiative decay widths of mesons and baryons are powerful tools for
understanding the structure of elementary particles and for constructing
dynamical theories of hadronic systems. Straightforward predictions for
radiative widths make possible the direct comparison of experimental data
and theory.  The small value of branching ratios of radiative decays makes
them difficult to measure directly, because of the large background from
strong decays.  Studying the inverse reaction
\mbox{$\gamma+\pi^-\rightarrow M^-$} provides a relatively clean method
for the determination of the radiative widths. Very good tracking 
resolution is needed and available in COMPASS to 
measure initial and final state momenta, and to thus  
exhibit the Primakoff signal at small four momentum
transfer $t$, where the
electromagnetic processes dominate over the strong interaction. 

We will study via COMPASS radiative transitions of incident
mesons to higher excited states. We will obtain new data \cite {hadron1} 
for radiative transitions leading from the pion to the $\rho$,
a$_1$(1260), and a$_2$(1320); and for the kaon to K$^*$ and higher
resonances.  Independent and higher precision data for these and higher
resonances will be valuable in order to allow a more meaningful comparison
with theoretical predictions. For example, the $\rho \rightarrow \pi
\gamma$ width measurements \cite{jens} range from 60 to 81 keV; the
a$_1$(1260) $\rightarrow \pi \gamma$ width measurement \cite{ziel} is
$0.64 \pm 0.25 $ MeV;  the a$_2$(1320)  $\rightarrow \pi \gamma$ width
\cite{ciha} is $\Gamma = 0.30 \pm$ 0.06 MeV. For K$^* \rightarrow K
\gamma$, the widths obtained previously \cite {berg,chan} ranged from
48-51 keV with errors 5-11 keV. 

\vspace{.5cm}
\noindent
{\bf The COMPASS Primakoff Trigger\\}
%\vspace{.5cm}

\noindent

For all the Primakoff physics
topics discussed above, due to the small
scattering angles, a trigger based on downstream information is
desired. The trigger should use the
characteristic pattern of a gently deflected hadron and one or two photon
hits in the EM calorimeter close to the neutral beam direction.
We design the COMPASS Primakoff trigger in this
way, which enhances the acceptance and statistics, and also
yields a trigger rate closer to the natural rate given by the low
Primakoff cross section.  We develop the trigger scheme with the help of
simplified MC generators, as follows:  a solution based on
fast scintillator hodoscopes that follow the hadron trajectory, and energy
deposit signals from photon calorimeters.

For COMPASS we forsee a three-level Primakoff trigger
scheme: T0 = beam definition, T1 = event topology, T2 = online software
filter. T0 is a fast logical relation between 1 mip 
(minimum ionizing particle) signals from thin
transmission and veto (hole)  scintillators before the target, and 
from the
upstream CEDARS Cherenkov counters providing beam PID. T1 is a downstream
coincidence between scintillation hodoscope signals from the pion track
and
total energy
deposit signal in the photon calorimeter. T2 is an
intelligent
software filter, placed in the DAQ stream after the event builder, which
counts the number of reconstructed segments downstream from the target, and
also sets cuts on event quality characteristics. Triggers for different
Primakoff topics may have different logic, but all will run in parallel.  

\begin{figure}[tbc]
\centerline{\epsfig{file=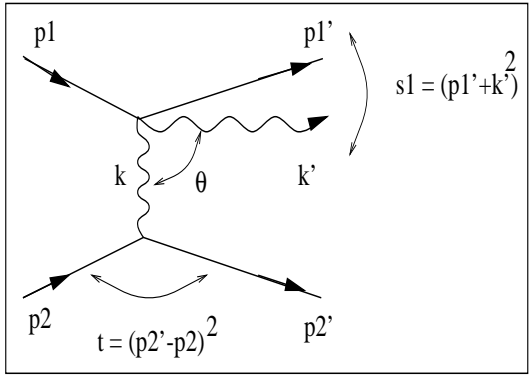,width=7cm,height=5cm}}

\caption{The Primakoff $\gamma$-hadron Compton process and kinematic
variables (4-momenta): p1, p1$^\prime$ = for initial/final hadron, p2,
p2$^\prime$ = for initial/final target, k, k$^\prime$ = for initial/final
gamma, and $\theta$ the scattering angle of the $\gamma$ in the alab
frame.}

\label{fig:diagram}
\end{figure}

As an example, pion polarizability events have a stiff pion at 
angles smaller than  0.5 mrad, 
very close to the non-interacting beam, and a single forward
photon at angles smaller than 
2 mrad. The trigger design should suppress the beam
rate
by at least a  factor 10$^4$, achieving high acceptance efficiency 
for Primakoff scattering events. 
The
kinematic variables for the pion polarizability
Primakoff process 
are shown in Fig.~\ref{fig:diagram}. A virtual photon from the 
Coulomb field 
of the target
nucleus 
is scattered from the pion and emerges as a real
photon accompaning the pion at small forward angles in the laboratory 
frame,
while the target nucleus (in
the
ground state)
recoils with a small transverse kick p$_t$. The peak at small  
target $p_t$ used to identify the 
Primakoff process is precisely measured
offline using the beam and vertex silicon detectors.

\begin{figure}[tbc] 
\centerline{\epsfig{file=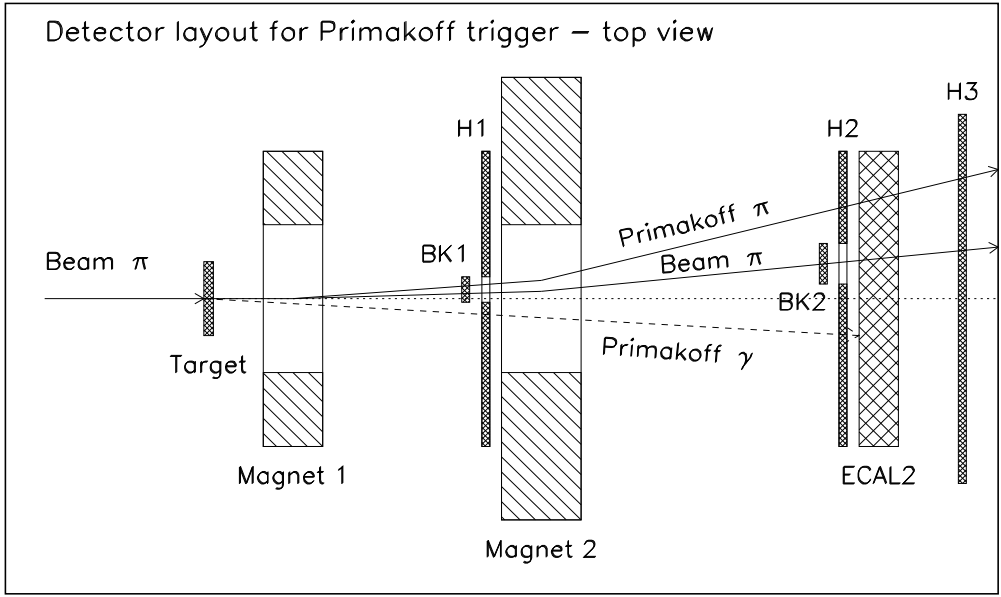,width=16cm,height=11cm}}
 
\caption{Detector layout for the COMPASS Primakoff trigger. BK1,BK2=beam
killer system, H1,H2,H3=hodoscope system for charged particle vetoing and
Primakoff pion detection, ECAL2=second photon calorimeter.}

\label{fig:trigger}
\end{figure}

The T1 trigger scheme (for photon-hadron reactions 
with one charged hadron and one or more photons in the final state)
is shown in Fig.~\ref{fig:trigger}. BK1 and BK2 
are  assumed to be small scintillator hodoscopes with 
moderate segmentation ($\sim$1cm) in the beam bend plane. 
 An anticoincidence on certain regions of the BK1/BK2
correlation 
 will serve to veto  non-interacting beam pions.
 According to Monte Carlo simulation, this achieves 
beam suppression (190 GeV/c) with 99.8\% efficiency, while maintaining 
100\%
 efficiency for all Primakoff scattered pions with momenta 
lower than 160 GeV/c.
 The BK1 hodoscope size should be optimized to accept both Primakoff
 and beam pions. The H1 hodoscope vetos
 charged particles with larger angles, and also events with higher multiplicity.
 The H2 and H3 hodoscopes could serve to veto non-Primakoff
 events and to also trigger on Primakoff scattered pions. 
This is achieved by measuring
the pion energy loss via characteristic angular
 deflection correlations in these hodoscopes. A  fast matrix chip
is needed for this purpose, as developed for the standard muon 
energy loss trigger in planned COMPASS studies of gluon polarization
in the proton.  
 In principle, a coincidence based on BK1/BK2 segmentated hodoscope
correlations, accepting correlation zones differing from the beam topology, 
could be also
 used to trigger on the Primakoff scattered pion.     

That trigger configuration strongly suppresses 
(with minimal acceptance loss)
backgrounds
associated with both non-interacting
beam particles and those involving nuclear interation of pions in the target
 and in COMPASS apparatus material.
 Finally, a decision signal for Primakoff pion detection,
 based on the hodoscope correlations, 
is required to be in coincidence with
a decision  signal for a minimum total energy produced by a Primakoff 
gamma-ray  
 in the electromagnetic calorimeter ECAL2.
The threshold for the ECAL2 signal needs to be optimized in order
 to affect only the kinematic region of 
the highest energy Primakoff pions, whose detection
 efficiency is already significantly reduced by the BK1/BK2 beam killer.

\begin{figure}[tbc] 
\centerline{\epsfig{file=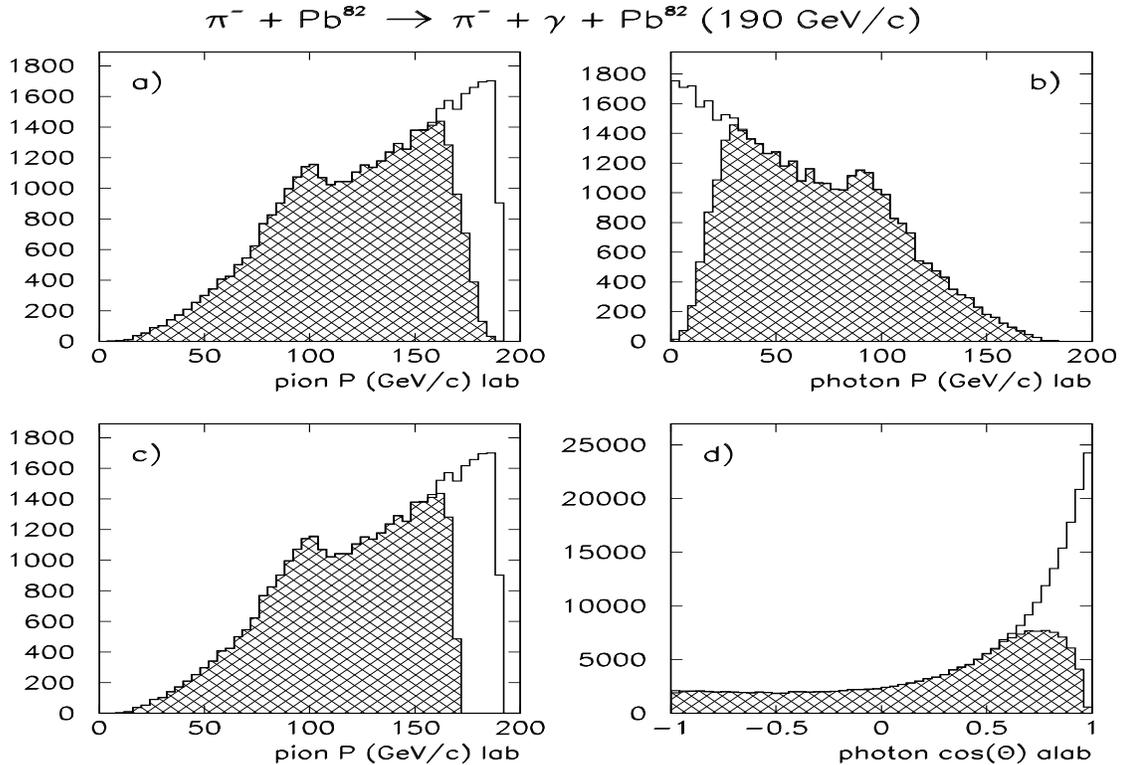,width=16cm,height=11cm}} 

\caption{MC simulation of the pion polarizability measurement in the
COMPASS 190 GeV/c beam: (a), (b) - Effect of beam killer system on the
acceptance of the Primakoff pion and photon momentum distributions; (c) -
Effect on Primakoff pion momenta distribution of beam killer system
together with ECAL2 set at 20 GeV lower threshold; (d) - Acceptance for
$\gamma\pi\rightarrow\gamma\pi$ angular distribution versus
$\cos(\theta)$, with $\theta$ the $\gamma$ scattering angle in the alab
frame. The hatched areas of the histograms correspond to kinematic regions
accepted by the trigger.}

\label{fig:acceptance} 
\end{figure}

We studied the acceptance for this trigger using our MC code POLARIS,
which generates Primakoff pion-photon (polarizability) 
interactions, with realistic beam
phase space. In Fig.~\ref{fig:acceptance} we plot
 the acceptance for the Primakoff pion and photon momenta distributions
 in the lab system,
 and for photon Compton scattering angle in the projectile (alab) frame. 
 The simulation was done for beam momentum of 190 GeV/c. Using the beam
 killer system maintains good acceptance for Primakoff pions
 of momenta $<$ 160 GeV/c and photons of momenta $>$ 30 GeV/c
 (see Fig.~\ref{fig:acceptance}a,b).
 Introducing the lower threshold for the ECAL2 signal to be equal 20 GeV,
 helps to suppress background processes, but does not affect
 acceptance in its most efficient region (see Fig.~\ref{fig:acceptance}c).
 Finally, for the given trigger design, we achieve
a large and flat acceptance versus
photon Compton scattering angles (see Fig.~\ref{fig:acceptance}d). This
is important to 
extract reliably the pion polarizability by a fit 
of the data to the theoretical
cross section.  
This trigger
does not affect the acceptance at the important back-angles where
the polarizability
contribution is largest. 

\vspace{0.5cm}
\begin{center}
{\bf  2. Plan of Operation:\\}
\end{center}
%\vspace{-.1cm}

\noindent
The CERN COMPASS experiment uses 50-280 GeV beams ($\mu$,
$\pi$, K, p)  and a virtual photon target, and magnetic
spectrometers and calorimeters to measure the complete kinematics of
hadron-photon reactions. 
The COMPASS experiment is currently under construction, and scheduled in
2000 to begin data runs, with muons initially and hadrons following.  The
Primakoff program in COMPASS is approved as the first hadron physics
program. The Primakoff physics program in COMPASS will
benefit from high statistics, excellent beam focussing and
momentum
analysis, and dedicated runs.

We need to achieve high statistics data with low systematic uncertainties
in this program. Accurate simulations are required using the new
C-programmed COMPASS COMGEANT package, presently being developped,
containing the updated experimental setup, trigger and the DAQ schemes,
accurate magnetic field mapping, and event reconstuction. We work on
this package and incorporate our Primakoff generators, including 
Hybrid Meson and Chiral Anomaly generators. 
We develop COMPASS event reconstruction
algorithms, and test them on COMGEANT simulated events. 

For the COMPASS Primakoff physics effort, we need to plan,
construct, and implement all hardware and software for all triggers of
interest for this physics. 
We prepare the COMPASS hadron-photon Primakoff trigger system by the
following phases: (1) further investigate trigger schemes, such as 
the
hodoscope matrix energy loss trigger discussed above, 
(2) refine our MC trigger simulations using COMGEANT, (3) construct the
trigger hardware, including an upgrade of the existing CEDARS Cherenkov
beam PID detector, scintillation hodoscopes, fast signal summing
circuitry,
mechanical supports, etc., (4) installation of the system at CERN, (5)
setting up the trigger detectors and electronics in the COMPASS muon beam,
(6) taking preliminary data with muons, writing event reconstruction
algorithms, and checking trigger performance, (7) use it for running the
COMPASS experiment with hadron beam.

%\newpage
\begin{center}
{\bf  3. Objectives and Expected Significance:\\}
\end{center}
%\vspace{-.1cm}

\noindent
The experimental pion polarizability determination to date has large
uncertainties. Kaon polarizabilities have never been measured. 
We will determine the $\gamma\pi$, $\gamma$K, 
and $\gamma$p
Compton
cross sections, and associated polarizabilities, and also the $\gamma\mu$ 
Compton scattering cross section as a check. 
For kaon studies, COMPASS results will be the first ever. For pion
studies, we will improve previous results by more than two orders of
magnitude, as we discuss below. 
The proton studies will provide a check on our methodology, since the
proton polarizability has been measured previously by standard $\gamma$p
scattering.

We studied the statistics attainable and uncertainties achievable for the
pion polarizabilities in the COMPASS experiment, based on Monte Carlo
simulations. 
We estimated 80 events/spill from the pion Primakoff effect,
corresponding to $10^{7}$ events per month at 100\% efficiency. We assume
a trigger efficiency of 75\%, an accelerator
operating efficiency of 50\%, and a tracking efficiency of 80\%. One may
then expect to observe as many as $3 \times 10^{6}$ Primakoff Compton
events per month of operation, following setup of COMPASS. Statistics of
this order will allow systematic studies, with fits carried out for
different regions of photon energy $\omega$, Z$^2$, etc.; and
polarizability
determinations with statistical uncertainties of order 0.2. For the kaon
polarizability, due to the lower beam intensity, the statistics will be
roughly 50 times lower. A precision kaon polarizability measurement
requires more data taking time, and should be carried out following the
analysis of the initial lower statistics data runs. 

Comparing chiral anomaly to polarizability data, we expect roughly 300 times 
lower statistics, due to the 140 times lower cross section and the lower 
$\pi^0$ detection efficiency \cite {ca}.

COMPASS can contribute significantly to the further investigation of
hybrids by studying Primakoff production of J$^{PC}= 1^{-+}$ $\rhot$
hybrids. The possibilities for Primakoff production of the $\rhot$ with
energetic pion beams, and detection via different decay channels has been
discussed by Zielinski et al. \cite {zihy}, and Monte Carlo simulations
for this physics were also carried out for 
COMPASS \cite {mpihy}. Considering vector dominance models, if
the $\rhot$ has a 1-10 MeV
branching width into the $\pi\rho$ channel, a branching width of $\rhot$
into the $\pi\gamma$ channel should be 3-30 keV \cite {zihy}. A hybrid
state with such a large radiative width would be produced at detectable
levels through the Primakoff mechanism in COMPASS. The COMPASS
trigger should allow observation of the $\rhot$ via the $\eta\pi^-$ decay
mode. With a relative P wave (L=1), the $\eta\pi^-$ system has J$^{PC}=
1^{-+}$. The other decay channels of $\rhot$ may be studied simultaneously
in COMPASS by relatively simple particle multiplicity triggers 
(three charged particles in final state, etc.). 

The evidence presented for the hybrid (pionic) meson offers COMPASS an
exceptional opportunity to take the next steps in this exciting field.
COMPASS can study hybrid meson candidates near 1.4 GeV produced by the
Primakoff process. COMPASS should also be sensitive to pionic hybrids at
higher excitation, and also to kaonic hybrids, which have not yet been
reported. We may obtain superior statistics for a hybrid state if it
exists, and via a different production mechanism without possible
complication by hadronic final state interactions. We may also get
important data on the different decay modes for this state. The
observation of this hybrid in different decay modes and in a different
experiment would constitute the next important step following the evidence
so far reported.

COMPASS can provide a unique opportunity to investigate QCD exotics,
glueballs and hybrids, produced via different production mechanisms:
central production for glueballs (not the subject of this proposal) 
and Primakoff production for hybrids.
Taking into account the very high beam intensity, fast data acquisition,
high acceptance and good resolution of the COMPASS setup, one can expect
from COMPASS the highest statistics and a "systematics-free" data sample
that includes many tests to control possible systematic errors. The
COMPASS effort should significantly improve our understanding of hybrid
physics. 

%\newpage
\begin{center}
{\bf  4. Acknowledgements:\\}
\end{center}
%\vspace{-.1cm}

The Tel Aviv U. group acknowledges support by the U.S.-Israel Binational
Science Foundation (BSF) and the Israel Science Foundation founded by the
Israel Academy of Sciences and Humanities. Thanks are due to B. Povh and J.
Pochodzalla of MPI Heidelberg (Kernphysik), for their hospitality and
interesting discussions during the preparation of this paper. Thanks are due
also to our COMPASS collaborators for continued interest: B. Barnett, F.
Bradamante, A. Bravar, S. U. Chung, M. Faessler, O. Gavrishuk, M. Lamanna, J.
Lichtenstadt, A. Olshevski, S. Paul, I. Savin, L. Schmitt, H.-W. Siebert, A.
Singovsky, V. Sougonyaev, G. Mallot, V. Poliakov, D. von Harrach, Th. Walcher,
U. Wiedner.

\end{document}